# ROOM AND CRYOGENIC TEMPERATURE BEHAVIOUR OF MAGNETIC SENSORS BASED ON GAN/SI SINGLE SAW RESONATORS

*Alexandra Nicoloiu[1], Florin Ciubotaru[2], Claudia Nastase[1], Adrian Dinescu[1], Sergiu Iordanescu[1], Hasnain Ahmad[2], Philipp Pirro[3], Christoph Adelmann[2] and Alexandru Müller[1]*
[1]IMT-Bucharest, ROMANIA,
[2]IMEC Leuven, BELGIUM and
[3]TU Kaiserslautern, GERMANY

## ABSTRACT

This work analyzes resonance frequency shift vs. the applied magnetic field strength for GHz operating GaN/Si SAW single resonators. Magnetostrictive elements (Ni and CoFeB) were deposited in the proximity of the interdigitated transducers (IDTs) of the resonators (A-type structures) and also over the IDTs, after covering them with a BCB layer to avoid short circuits with IDTs metal (B-type structures). This work targets emerging applications of SAW resonators in driving spin wave pumping and in coupling of surface acoustic waves (SAW) with superconducting Q-bits. Magnetic sensitivity of the SAWs was analyzed at room temperature (RT) and at cryogenic temperatures, obtaining high magnetic sensitivities at 16 K. According to our knowledge, GaN based SAWs are first time used in magnetic applications; also, cryogenic behavior of magnetic SAW sensors is first time analyzed.

## KEYWORDS

magnetic field, SAW sensor, cryogenic measurements, GaN/Si.

## INTRODUCTION

Rayleigh waves propagation in thin magnetostrictive films deposited on piezoelectric substrates was studied in the past. A new interest in this topic appeared due to the possibility to drive spin pumping with SAW resonators [1-3]. Recent experiments have demonstrated that the propagation of SAWs in piezoelectric substrates can be controlled by a thin magnetic layer placed on the substrate [4-7].

High performance magnetic SAW type sensors using magnetostrictive layers (e.g. for the IDTs) were also targeted in recent studies [6]. One-port SAW magnetic sensors having grooved Ni electrodes on ST-90°X quartz showed higher sensitivities than the conventional SAW structures with Ni electrodes on the same substrate [8]. A layered SAW structure (Ni/ZnO/IDT/LiNbO3) was simulated and optimized in terms of its sensitivity to the magnetic field intensity in [9]. Moreover, the authors showed through simulations that substituting ZnO with alumina, the sensor sensitivity increases 9 times in X direction.

The developed SAW devices presented in literature are based on bulk piezoelectric materials (LiNbO$_3$, quartz, etc.) where resonance frequencies can't exceed 2.5 GHz. Such frequencies are too low for spin wave excitation in typical ferromagnets with high magnetostrictive response, therefore, such devices are forced to operate at high harmonics, which is not ideal for magnetoelectric devices.

Recently, GaN/Si SAW devices operating at frequencies above 5 GHz were used as temperature and pressure sensors [10, 11, 12, 13].

In the present work, magnetic sensors based on SAW devices manufactured on GaN/Si, are developed. The high operating targeted resonance frequency (6-12 GHz) and the low associated dimension will facilitate their potential applications to drive spin pumping.

Their magnetic sensitivity was analyzed not only at room temperature but also at very low temperatures (< 25 K) which can be significant also for emerging applications connected to SAW coupling with superconducting Q-bits [14].

## MANUFACTURING

To allow the propagation of the SAW wave in the proximity of the IDTs, single resonator structures, without reflectors, have been fabricated using advanced nanolithography, similar to those presented in [10, 11]. GaN/Si wafers, obtained on a commercial basis from NTT-AT, Japan, have been used. A buffer layer with a total thickness of 0.3 μm was grown between the high resistivity Silicon substrate (500 μm thick) and the 1 μm thin, undoped GaN top layer.

First, the coplanar waveguide ports were patterned for on-wafer $S_{11}$ parameter measurements. For this, conventional photolithography, e-gun metallization (Ti/Au 20/200 nm) and lift-off technique have been used. Electron Beam Lithography was used for the IDT structure (150 fingers and interdigit spacings 170 nm wide). The IDTs were made with Ti/Au (5 nm/ 95 nm metallization thickness), deposited by e-gun evaporation and selectively removed by lift-off process.

Different configurations of magnetostrictive strips were developed on one or both sides of the IDTs. For the A-type structures, Ni strips, 100 nm thin, were deposited in the proximity of the IDTs. The CoFeB strips (18 nm thin) were deposited also by e-beam and lift-off techniques.

For the B-type structures, where the magnetostrictive layer is deposited over the IDTs, a dielectric layer was first deposited over the IDTs, to avoid a short circuit between the metal of the IDTs with the magnetostrictive layer. On these devices, strips of magnetostrictive layers are also deposited in the proximity of the IDTs. The SAW devices have similar topologies with the A-type structures.

A SEM photo of a manufactured A-type structure is presented in Fig. 1.

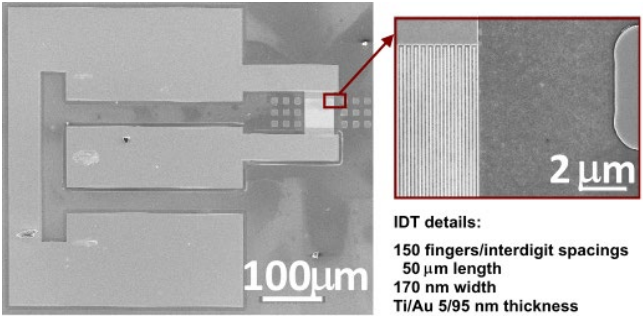

*Figure 1: SEM photo of an A-type SAW structure. Inset: detail of the IDT area.*

## EXPERIMENTAL RESULTS

The sensor was assembled on a fixture inserted in a cryostat and it was connected to a Vector Network Analyzer through probe tips (150 μm pitch), allowing S-parameter measurements. The cryostat is placed between the poles of an electromagnet able to provide maximum 4200 Oe at a distance suitable for the cryostat dimensions.

The resonance frequency at H = 0 is 6.31 GHz. Figures 2 and 3 present the influence of the magnetic field, H, on $S_{11}$ parameter measured for the A-type structure, with CoFeB strips, at room temperature and respectively, at 16 K. The structures should be precisely maintained at fixed temperature when they are measured at RT, because around the room temperature, the resonance frequency is very sensitive to temperature changes [10], in contrast to their behavior at low temperatures (< 25 K), where also the magnetic sensitivity is much higher.

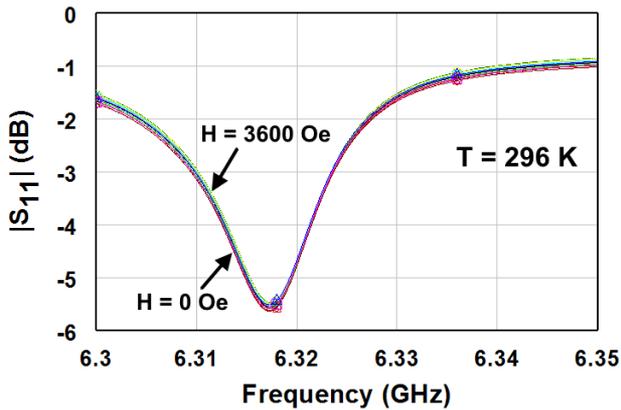

*Figure 2: The influence of H on the $S_{11}$ parameter measured at room temperature.*

The resonance frequency shift vs. magnetic field strength was measured in a cryostat fixture, at RT and at temperatures below 25 K, for H = 0…3600 Oe.

For the **A-type structures** (Fig. 4 - Ni strips, Fig. 5 – Ni strip and Fig. 6 - CoFeB strip with identical geometries), at room temperature, the relative frequency shift slowly increases to values up to 25 ppm for Ni and to about 40 ppm for CoFeB striped structures.

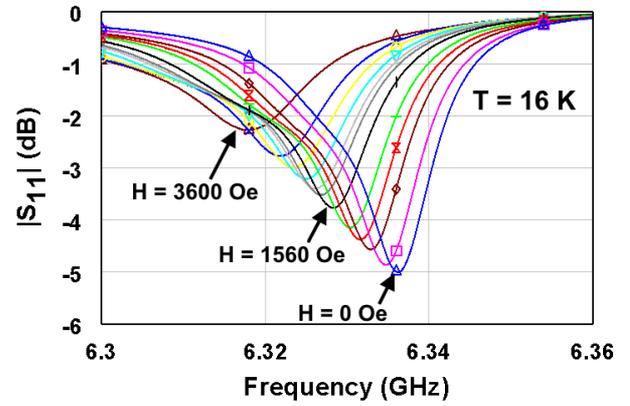

*Figure 3: The influence of H on the $S_{11}$ parameter measured at 16 K.*

At 16 K the relative frequency shift is negative, almost one order of magnitude higher for CoFeB than for the Ni striped devices.

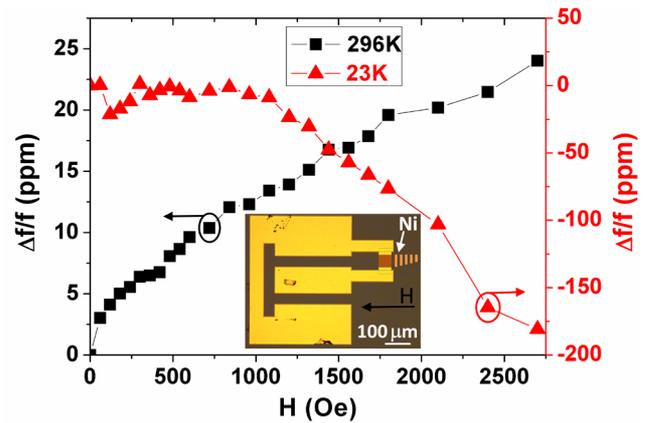

*Figure 4: Relative frequency shift vs. H for SAW structure A-type with Ni strips in the proximity of IDT.*

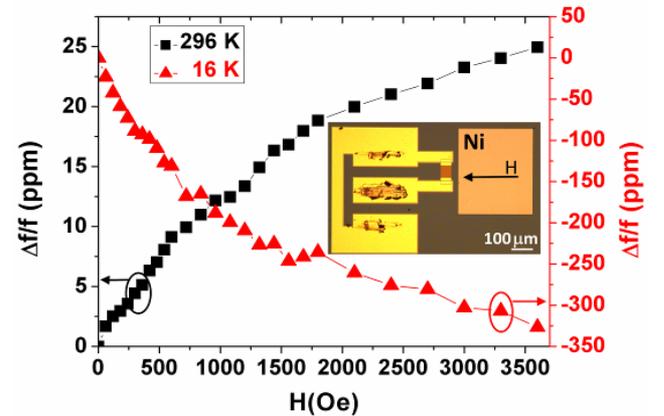

*Figure 5: Relative frequency shift vs. H for SAW structure A-type with Ni layer in the proximity of IDT.*

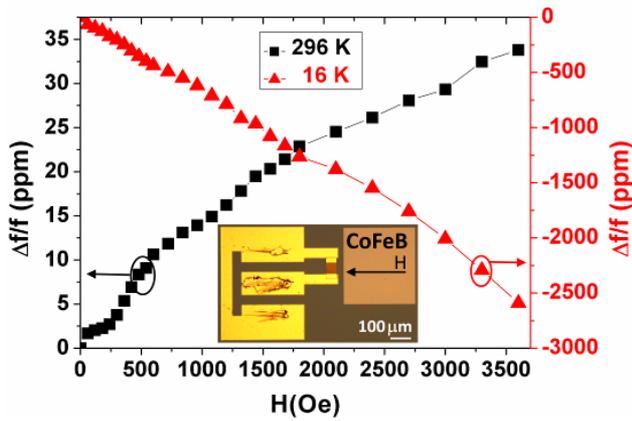

*Figure 6: Relative frequency shift vs. H for SAW structure A-type with CoFeB layer in the proximity of IDT.*

The experimental results showing the relative frequency shift vs. H for the **B-type structures** (Ni strip over the IDTs and Ni lateral strips with different configurations) are presented in Fig. 7 and Fig. 8.

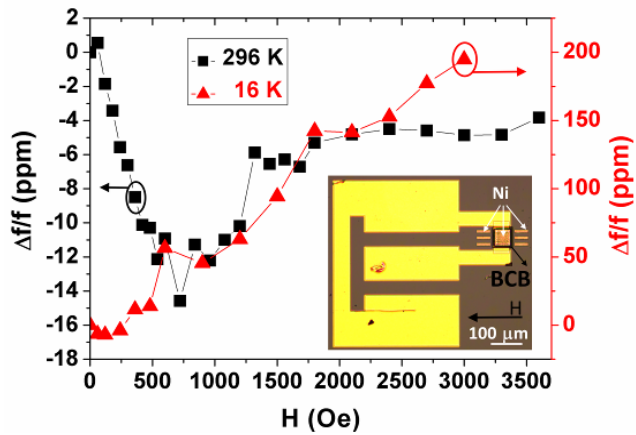

*Figure 7: Relative frequency shift vs. H for SAW structure B-type, with Ni strips over and in the proximity of IDT (first configuration). A BCB layer is deposited prior to Ni over the IDTs in order to avoid short circuits between the IDTs metal and Ni.*

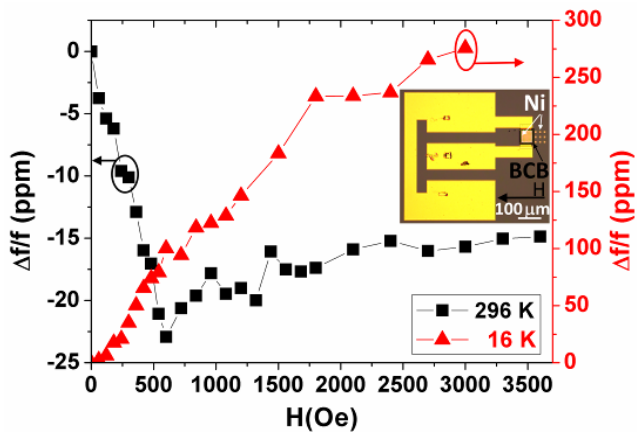

*Figure 8: Relative frequency shift vs. H for SAW structure B-type, with Ni strips over and in the proximity of IDT (second configuration).*

The behavior is strongly nonlinear at RT, with saturation starting below H = 1000 Oe and low magnetic sensitivities. At 16 K the effect of the magnetic field becomes monotone and the relative frequency shift increases at values beyond +200 ppm. This behavior could be linked to:

(i) the shrinking of the digit/interdigit spacing when the magnetostrictive strip is placed atop (having as effect the increase of the resonance frequency) and its expansion when the strip is placed sideways (having as effect the decrease of the resonance frequency);

(ii) the much higher magnetostriction at low temperatures for both materials and its strong nonlinearity vs. H;

(iii) the comparable effect of sound velocity variation in magnetic field with the digit/interdigit spacing dimension change at RT.

The results of Δf/f are referenced to a structure without magnetostrictive layer, which showed negligible change of the resonance frequency with the magnetic field variation.

An efficient way to increase the frequency shift vs H, at RT, is to deposit the magnetostrictive layer directly over the IDT metallization. Preliminary results confirmed an increase of about 3 times.

## CONCLUSIONS

The influence of lateral magnetostrictive strips on the change of the resonance frequency vs. magnetic field strength for GaN/Si SAW structures was analyzed. Measurements were performed at room temperature and at cryogenic temperatures in the 16-25 K range. The magnetostrictive effect is more than one order of magnitude higher for cryogenic temperatures compared to RT. These values are influenced by the higher magnetostriction at low temperatures and the sound velocity variation vs. the digit/interdigit spacing dimension change for various shapes and positions of the magnetostrictive layer.

Future plans include the development of SAW structures with IDTs Ti/Au/CoFeB, having also lateral CoFeB strips.

## ACKNOWLEDGEMENTS

The authors acknowledge the support of the EU under H2020 FET Open Project CHIRON (grant agreement no 801055). The Romanian authors also acknowledge the Ministry of Research and Innovation, CNCS - UEFISCDI for the support of this work under the project SupraGaN PN-III-P4-ID-PCE-2016-0803 within PNCDI III.

## CONTACT

*Alexandra Nicoloiu, tel: +40212690775; alexandra.nicoloiu@imt.ro

*Alexandru Müller, tel: +40212690775; alexandru.muller@imt.ro